# Voter coalitions and democracy in Decentralized Finance: Evidence from MakerDAO[1]

**Xiaotong Sun[1]**

**Xi Chen[2]**

**Charalampos Stasinakis[3]**

**Georgios Sermpinis[4]**

[1]Adam Smith Business School, University of Glasgow, Gilbert Scott Building, Glasgow G12 8QQ, United Kingdom. Email: Xiaotong.Sun@glasgow.ac.uk.

[2]Leonard N. Stern School of Business, New York University, Kaufman Management Center, New York, NY 10012, United States. Email: xc13@stern.nyu.edu

[3]Adam Smith Business School, University of Glasgow, Gilbert Scott Building, Glasgow G12 8QQ, United Kingdom. Email: Charalampos.Stasinakis@glasgow.ac.uk,

[4]Adam Smith Business School University of Glasgow, Gilbert Scott Building, Glasgow G12 8QQ, United Kingdom. Email: Georgios.Sermpinis@glasgow.ac.uk.

**Abstract**

Decentralized Autonomous Organization (DAO) provides a decentralized governance solution through blockchain, where decision-making process relies on on-chain voting and follows majority rule. This paper focuses on MakerDAO, and we find three voter coalitions after applying clustering algorithm to voting history. The emergence of a dominant voter coalition is a signal of governance centralization in DAO, and voter coalitions have complicated influence on Maker protocol, which is governed by MakerDAO. This paper presents empirical evidence of multicoalition democracy in DAO and further contributes to the contemporary debate on whether decentralized governance is possible.

---

[1] The online appendices can be accessed: https://drive.google.com/file/d/1Xbgq7S73sV3B2Gv_vfLzb2bzAWmiHWZ7/view?usp=sharing

# 1.Introduction

Decentralized Finance (DeFi) encodes the logic of traditional financial systems and benefits from characteristics inherited from the underlying blockchain infrastructure. However, decentralization, as the most controversial virtue of DeFi, is often questioned. Recently, Carapella et al. (2022) argue that DeFi is not immune from centralized governance, and associate problems may arise.

The most popular mode of governance in DeFi is *Decentralized Autonomous Organization (DAO)*, proposed by Jentzsch (2016). In DAO, there is no centralized coalition, and any suggested changes to DAO should be jointly decided by DAO members. To distribute decision-making power, DAO will usually issue *governance token*, which are tradable cryptocurrency. Voters' decision-making power relies on the amount of governance token, and the proposal that gets most voting power will be implemented. Benefiting from the transparency and accuracy of blockchain, voting results are publicly visible and hard to be tampered, and DAO has been the most widely adopted choice for on-chain governance.

Unfortunately, governance centralization is an inevitable problem for DAO. First, blockchain itself is not safe haven for decentralization. Buterin (2021), as the co-founder of Ethereum blockchain, proposes that complete decentralization is impossible if blockchain pursues scalability and securities at the same time. In fact, a few key developers propose most changes to blockchain (Hsieh et al., 2017; Yermack, 2017), implying that these developers have more control. Furthermore, DAO is not immune from governance centralization. By investigating voting history of several leading DAO, DAO governance is controlled by several dominant voters, and their centralized power has complicated influence (Sun, Stasinakis and Sermpinis, 2022; Fritsch, Müller and Wattenhofer, 2022;). For example, Carter and Jeng (2021) contend that key decision-makers, especially administrative teams can abuse their governance power. Using *Decentraland* as a case study, Goldberg and Schär (2023) prove that centralized governance may result in rent extraction behavior and other problems. Moreover, Nadler and Schär (2020) show that some DAO participants may conceal their centralized decision-making power by creating multiple on-chain identities.

The previous studies mainly focus on centralization problems at individual level, however, DAO governance is likely to be battlefield of voter coalitions with different interests. Given the fact that voting is all about social choice functions (Arrow, Sen and Suzumura, 2011; Kelly, 1988; Plott, 1976; Schwartz, 1986), voters with similar interests and characteristics can form coalitions (Downs, 1957; Black, 1990; Enelow and Hinich, 1994; Tajfel and Turner, 2004), while voters with different beliefs will vote against each other (Adams, Merrill and Grofman, 2005; Abramson et al., 2009). In corporate finance, voting rights are distributed among shareholders, and both small shareholders and large shareholders can form their own coalitions. Smaller shareholders collaborate in order to protect their own rights (Bennedsen and Wolfenzon, 2000; Zwiebel, 1995), while large shareholders attempt to extract private benefits by seizing more control (Bennedsen and Wolfenzon, 2000; Dyck and Zingales,

2004). In DAO, we expect to detect voter coalitions since both dominant voters and minority voters exist, and the interlinks between voters can help to depict DAO governance better.

This paper attempts to answer two key research questions: (1) if voter coalitions exist in DAO; (2) how voter coalitions influence DAO. We choose the most influential DAO, namely MakerDAO, as a case study. MakerDAO is probably the most successful DAO in blockchain because it sets industry standards, e.g., 'one token – one vote' principle and a combination of on-chain governance and off-chain discussion. The main role of MakerDAO is to manage Maker protocol, which is a leading financial system on Ethereum blockchain. Maker protocol issues Dai (DAI) stablecoin, which is soft-pegged to US dollar, and any agents can borrow DAI by locking collateral. In other words, Maker protocol is a multi-collateral DAI system, where MakerDAO decides how this system develops.

To study MakerDAO, we retrieve voting history of governance polls from 15$^{th}$ August 2019 to deployed on 25$^{th}$ July 2022, where all voters' choices and voting power are available. After applying K-means clustering algorithm to MakerDAO dataset, we identify three distinguished voter coalitions (i.e., coalitions 1-3), and the largest coalition, mentioned as coalition 1, has the most voters and contribute to most total votes. Though the other four coalitions have only a few voters, they can win governance polls of Maker protocol, implying that MakerDAO governance is battlefield for these coalitions.

By applying a series of factor analysis, we show that the voter coalitions have dissimilar effects on Maker protocol. The first sub-section of empirical results centers on the coalitions' voting share in every governance poll. Here, three factors specific to Maker protocol are studied, including DAI volatility, daily revenue and new users of Maker protocol. When voting share of the largest coalition is higher, DAI will be more volatile, and daily revenue of Maker protocol will decrease. However, smaller voter coalitions have the opposite effects. Previous literature (e.g., Sah and Stiglitz, 1988; Sah and Stiglitz, 1991) mainly discuss how (de)centralization affects firm performance, however, they ignore that coalitions may affect the underlying financial system very differently.

Beside voting share, we are interested in if group cohesion also matters for voter coalitions, since previous studies point out that cohesive teams can contribute to better firm performance (Hogg, 1992; Pepitone and Reichling, 1955; Schachter et al., 1951). We adopt a measurement named *Agreement Index (AI)* (Hix, Noury and Roland, 2005), which is widely used in political science. A higher AI means that group members tend to choose the same option, while AI will be close to 0 if voting power of a group is equally distributed among all options. Again, the largest coalition has different effects compared to the smaller coalitions. For example, more new users will be attracted to Maker protocol when the largest coalition is more cohesive, while cohesive smaller coalitions can decrease the number of daily new users.

We also consider the interlinks between Maker protocol and crypto market. DAI, as a primary stablecoin, can be traded in various blockchain-based platforms. Here we choose five destinations of financial flows of DAI, including *Centralized Finance (CeFi)*, *Decentralized Exchanges (DEXes)*,

*Lending Protocols (LPs)*, *External Owned Address (EOA)*, and *bridges*[2]. Our findings show that voter coalitions can drive DAI flows in differently ways. For example, the voting share of a small voter coalition can increase the number of DAI transferred to CeFi. The findings imply that voters coalitions have certain similar preference, e.g., where to transfer DAI.

The remainder of our paper is organized as follows. Section 2 provides necessary background knowledge and related work. The dataset for MakerDAO, clustering algorithm, and measurements of group cohesion are defined in section 3. The main empirical results are presented in section 4, and section 5 concludes.

## 2. Background
### 2.1 Decentralized Finance (DeFi) and Decentralized Autonomous Organization (DAO)

Powered by programmable blockchain, any agents can replicate financial activities on blockchain, where the third coalition is not necessary component. Simply, *Decentralized Finance (DeFi)* refers to blockchain-based financial applications without any centralized intermediaries. DeFi can replicate most traditional financial systems, such as trading platforms and borrowing and lending marketplaces, and the rapid growth of DeFi brings forward both opportunities and challenges. For more details of DeFi and its potential risks, we refer readers to Werner et al. (2021), Makarov and Schoar (2022), and Carapella et al. (2022).

Beside financial risks inherently existed in DeFi, another inevitable plain point is: how to govern DeFi? Without relying on any third coalition, DeFi naturally tends to expand such decentralization to its governance. Among all novel solutions to decentralized governance, the most widely adopted organizational form is *Decentralized Autonomous Organization (DAO)*, which is first proposed by Jentzsch (2016). Formally, DAO is an entity structure lead by community instead of centralized authority. In a DAO, decision-making power is distributed among all DAO members, and decisions about the future of the DAO will be jointly made by members via voting. In other words, DAO is owned and managed by their members.

In practice, most DeFi protocols are governed by DAO (e.g., Maker protocol is governed by MakerDAO), and DAO will usually issue their own *governance token*, which resembles shares in corporate finance. If there are any suggested changes to the DeFi protocol, governance token holders can state their opinions via voting, and their voting power relies on the amount of governance token. Benefiting from underlying blockchain, voting records can be documented on blockchain, and on-chain voting is more precise and transparent than decision-making process in corporations (Hsieh, Vergne and Wang, 2017).

DAO governance relies on voting. Among all voting procedures, most DAO chooses Plurality voting procedure, meaning that only the candidate who gets votes more than any other counter-coalition will be elected (Arrow, Sen and Suzumura, 2011). In DAO, given a proposal, only the option that get the most votes will be deployed. As for voting principle, though quadratic voting exists, most DAOs adopts 'one token – one vote'. However, Burkart and Lee (2008) argue that such a voting mechanism will be optimal only when several agents compete.

---
[2] In the context of blockchain, bridges refer to tools that connect two blockchains and allow agents to transfer on-chain assets from one blockchain to the other.

**2.2 MakerDAO and Maker protocol**

Created in 2014, MakerDAO has grown up to the most successful DAO (MakerDAO, 2020). Governed by MakerDAO, Maker protocol has adopted large market share in on-chain lending. Maker protocol issues two native cryptocurrencies, namely Dai (DAI) and Maker (MKR). DAI is a stablecoin soft-pegged to the US dollar, and people can borrow DAI by locking collaterals (usually cryptocurrencies accepted by Maker protocol). In a way, Maker protocol is a Multi-Collateral Dai (MCD) system. MKR is the governance token, and any MKR holders can participate in on-chain voting in Maker protocol. For more details of the governance structure in MakerDAO, we refer readers to Sun, Stasinakis and Sermpinis (2022).

Currently, Maker protocol applies 'one token – one vote' rule, therefore, voters with more MKR will have more decision-making power. Empirically, a small group of voters control most voting power in MakerDAO, and governance centralization can be witnessed in other DAOs as well (Fritsch, Müller and Wattenhofer, 2022). Surprisingly, though governance centralization is not intention of DAO believers, centralized decision-making power can have positive effects on the underlying DeFi protocol (Sun, Stasinakis and Sermpinis, 2022). In this paper, we dive more deeply into the voting history of MakerDAO, and beside individual level of governance centralization, more information can be revealed by considering potential coalitions of voters.

**2.3 Governance centralization**

In traditional finance, governance centralization is regarded as a double-edge sword. A well-studied example is banks. Iannotta et al. (2007) prove that higher ownership concentration can contribute to better loan quality and lower insolvency risks. However, bank managers' dominant decision-making power can lead to more risk-taking decisions, and unacceptable outcomes can happen (Mollah and Liljeblom, 2016; Dbouk et al, 2020). Since Maker protocol resembles a bank in crypto markets, we suspect similar complex findings should be observed.

In corporate finance, governance centralization also attracts discussion (e.g., Shleifer and Vishny, 1997). Centralized ownership facilitates interest alignments (Shleifer and Vishny, 1997), but blockholders can have self-serving actions (Burkart and Lee, 2008). Adding to the problem, small stakeholders lack incentives to participate in corporate governance (Burkart and Lee, 2008). Currently,

DAOs suffer from similar problems, i.e., low participation of governance.

However, some literatures argue that decentralized governance is the source of uncertainty. An important reason is that decision makers will propose more choices (Allen and Gale, 1999), and thereby the decisions made by an organization will be lack of consistency (Garlappi et al., 2017). Moreover, Demsetz and Villalonga (2001) demonstrate that firm performance may be not statistically related to ownership structure.

Previous studies show very dissimilar conclusions, implying that the internal structure of governance has not been well explored. Here, a pain point is data limitation since the internal structure of corporate governance is not very transparent (e.g., Hermalin and Weisbach, 2003; Adam, Hermalin and Weisbach, 2010). However, in DAO, all voters and their voting patterns are publicly observable, and we can examine if arguments in corporate finance can apply to DAO and DeFi.

**2.4 Shareholder coalition**

Governance will be even more complicated when shareholder coalitions exist. From a formal economic perspective, the emergence of shareholder coalitions seems to be inevitable (Maury and Pajuste, 2005; Crespi and Renneboog, 2010). Usually, shareholders with similar goals can form coalitions (Black, 1990; Tajfel and Turner, 2004). For example, to influence corporate policy, minority shareholders can form a shareholder coalition (Zwiebel, 1995; Bennedsen and Wolfenzon, 2000). On the other hand, Bennedsen and Wolfenzon (2000) show that large shareholders can form coalitions as well, and several coalitions compete to seize control for their own private benefits. Driven by private benefits, shareholder coalitions may withhold critical information, though extremely negative actions, e.g., lying or stealing, are not very common (Wathne and Heide, 2000).

To defend private benefits, shareholder coalitions will attempt to affect vote outcomes. Large shareholders, e.g., institutional investors, tend to support management, and they may never vote against management-sponsored proposal (Dressler, 2020), and Matvos and Ostrovsky (2010) show that some shareholders are consistently friendly to management in voting process. However, conflicts of interest can destroy such harmony in voting process (Hamdani and Yafeh, 2013; Cvijanović et al., 2016), especially when it comes to controversial proposals (Ginzburg et al., 2022). For example, outside shareholders could vote against inside shareholders or management (Marquardt et al., 2018).

In DAO, the holders of governance token resemble shareholders in corporate finance. Chohan (2017) and Schär (2021) argue that groups of voters may attempt to manipulate the DAO governance. Goldberg and Schär (2023) present some preliminary evidence by using *Decentraland* as a case study. Large voters exist in governance of Decentraland, and they can collude with small voters to affect the outcomes of voting. For example, large voters can choose the timing of casting their vote, and other voters could be influenced by large voters' strategic behavior. Previous studies have discussed if voters can collude, while this paper will further explore if voters can form coalitions and thereby affect the underlying DeFi protocol. Given that all transactions are publicly observable, the clustering algorithm is feasible, and voters' private benefits can be better investigated by considering their voting patterns.

## 3. Voter coalitions in MakerDAO
### 3.1 Data Collection

The details of governance polls and voting history are publicly available. In *Maker Governance Portal*, poll details, including titles, review of proposals, and options can be found. For each poll, several labels, e.g., 'risk parameter' and 'collateral onboard' are shown for better understanding the content of governance polls. To get the voters' addresses, their choices and voting power, we query the voting history from *MCD Voting Tracker*. We investigate governance polls from Poll 16 (deployed on 15th August 2019) to Poll 838 (deployed on 25th July 2022). Poll 16 is the first governance poll that MKR holders can participate in. Some polls failed[3], so they are not documented in the portal. Hence, the dataset consists of a total of 809 successful governance polls. After retrieving voters' addresses, 1717 unique voters are found, and the voters' public names and their labels can be manually collected by searching for their addresses on Maker Governance Portal and *Watchers.pro*.

The first step of data pre-process is to replace textual options with numerical values. Most Maker governance polls have three options, including "Yes", "No", and "Abstain". For this type of polls, we will assign 1, -1, 0 to "Yes", "No", and "Abstain", respectively. For other polls, we also assign integer values to different options. Noticeably, in all governance polls, we assign 0 to "Abstain".

Another question is how to assign value if a voter (e.g., voter $i$) does not participate in a certain poll (e.g., poll $j$). In the context of clustering algorithms, such cases will bring forward missing values, i.e., *NA* in the dataset. Two common solutions are (1) to delete observations with NAs and (2) to fill NAs with the mean, however, these two solutions are not the best choices here. The first solution will delete voters who do not participate all voting polls, as a result, few voters will be preserved. The second solution will misinterpret the nature of not participating. If we fill NAs with mean, in a way, we 'make decisions' on behalf of the voters who do not vote. But their actions, i.e., not voting, imply that they abstain. Therefore, we will assign 0 to NAs, meaning that voter $i$ that does not participate in poll $j$ will choose "abstain".

### 3.2 Identification: Clustering and group cohesion

In order to alleviate the aforementioned issues, we need to apply a clustering algorithm. As a first step, we need to pre-process voting datasets, including data standardization and dimension reduction.

---
[3] Poll 28, 39, 47, 69, 78, 183, 282, 284, 286, 500, 604, 769, 818 and 821 failed.

Given a dataset $X$, the formula of transferring $x \in X$ is

$$\frac{x - \bar{X}}{X.\text{std}} \quad (1)$$

Where $\bar{X}$ is the mean of dataset $X$ and $X.\text{std}$ refers to the standard deviation.

In the context of clustering algorithm, each poll in voting datasets is a feature. With more than 800 polls, the dataset for voting history is high-dimensional, and we need to reduce dimensions for better modelling. Here, we choose *Principal Component Analysis (PCA)* for dimensionality reduction. Simply, PCA can compute the principal components of a dataset and only keep the first few ones. In this way, a high-dimensional dataset can be transferred to a lower-dimensional dataset without losing much of data's information. Generally, the new dataset generated by PCA should keep at least 95% of variance in the original dataset, therefore, we preserve 115 principal components and 95.01% of variance is contained in the lower-dimensional dataset.

To detect voters with similar voting patterns, we choose *K-means*, which is a widely adopted clustering algorithm. Given a set of voters' voting history $(v_1, v_2, \ldots, v_n)$, where each voting history is a $d$-dimensional real vector, K-means aims to cluster voters into $k (\leq n)$ sets $V = \{V_1, V_2, \ldots V_k\}$ so as to minimize the *within-cluster sum of squares (WCSS)*.

Formally, the objective is to find

$$\underset{V}{\text{argmin}} \sum_{i=1}^{k} \sum_{x \in S_i} ||v - \mu_i||^2 = \underset{V}{\text{argmin}} \sum_{i=1}^{k} |V_i| Var V_i \quad (2)$$

where $\mu_i$ is the mean of points in $V_i$.

In our case, $v_i$ is a vector that records voter $i$'s choices in all governance polls, and $d$ denotes the number of governance polls in the MakerDAO dataset. Given a poll $j$ and voter $i$, if the voter do not participate in poll $j$, then $v_{i,j} = 0$, which has the same value as "Abstain". Assuming that K-means can generate $k$ sets, i.e., $V = \{V_1, V_2, \ldots V_k\}$, each $V_i$ can be regarded as a 'voter coalition', where voters share similar voting patterns.

To run K-means, the number of clusters, i.e., the parameter $k$ should be optimally chosen. Two common criteria are *elbow method* and *silhouette score*, and more formal introduction can be found in Malik and Tuckfield (2019). Simply, an optimal cluster number, i.e., $k^*$, should have high silhouette score, and the curve of distortion score flattens when $k$ is larger than $k^*$. Combing information in Figure 1, we choose $k^* = 3$.

[Figure 1 here]

An important step for our analysis is to measure the voting group cohesion. Given a voter coalition, their members may have split opinions on certain governance polls. Intuitively, less division of opinions implies better group cohesion of a coalition. Here, we introduce the modified index of *Agreement Index (AI)*. Previously, Rice (1928) develops an index to measure the rate of 'not voting identically', however, this index can only describe 'yes' – 'no' option. Then, Hix, Noury and Roland (2005) introduce *Agreement Index (AI)*, which can be applied to polls with three options, i.e., "yes", "no", and "abstain". Formally, AI of voter coalition $i$ can be calculated as:

$$AI_i = \frac{\max\{Y_i, N_i, A_i\} - \frac{1}{2}[(Y_i + N_i + A_i) - \max\{Y_i, N_i, A_i\}]}{Y_i + N_i + A_i} \quad (3)$$

where $Y_i, N_i, A_i$ denote the number of "yes", "no" and "abstain" votes, respectively.

Similarly, we can expand AI to polls with $j (\geq 3)$ options as below:

$$AI_i = \frac{\max\{Option_1, \ldots, Option_j\} - \frac{1}{j-1}[(Option_1 + \cdots + Option_j) - \max\{Option_1, \ldots, Option_j\}]}{Option_1 + \cdots + Option_j} \quad (4)$$

where $option_j$ denote the number of votes of option $j$.

Given a voting poll and voter coalition $i$, $AI_i$ will be a numeric value between 0 and 1. A higher $AI_i$ means better group cohesion. For example, if all members of coalition $i$ choose the same option, $AI_i$ should equal to 1. However, if the votes of coalition $i$ are equally divided among all available choices, $AI_i$ will be 0.

### 3.3 Detection of voter coalitions in MakerDAO

Initially, our focus is on the descriptive statistics of both polls and voter coalitions. Table 1 provides descriptive statistics of Maker governance polls, specifically from Poll 16 to Poll 838. Additionally, we present the total votes and the number of voters in Figures 2 and 3, respectively. While the total votes show a gradual increase over time, the number of voters exhibits volatility. Most polls have fewer than 60 voters, which represents a small group in comparison to the total number of users within the Maker protocol. Consequently, the decision-making power is largely controlled by voters who frequently participate in voting and possess a significant MKR balance. If voter coalitions exist, the Maker governance system becomes an illustration of competition between coalitions on the blockchain.

**[Table 1 here] [Figures 2 and 3 here]**

Next, we proceed with the detection of voter coalitions using the K-means clustering algorithm. We divide the Maker governance polls into two subsets: Poll 16 (started on August 15th, 2019) to Poll 412 (started on January 11th, 2021), and Poll 413 (started on January 18th, 2021) to Poll 838 (started on July 22nd, 2022). To mitigate endogeneity concerns, we apply the K-means algorithm to the first subset and conduct factor analysis using the second subset. We exclude minority voters whose total votes are less than 500 MKR, resulting in 172 remaining voters. Notably, these voters have participated in at least 5 governance polls. Table 2 displays the results, revealing the presence of 3 voter coalitions. Voter coalition 1 comprises the largest number of members and significantly more total votes compared to the other two coalitions. Moreover, coalition 1 has participated in the majority of governance polls. On the other hand, although coalitions 2 and 3 are smaller in size, their total votes should not be disregarded.

**[Table 2 here]**

Figure 4 displays the voting share of the three identified coalitions. It is worth noting that the voting share of the other four coalitions also varies, and in certain polls, these four coalitions have accounted for a dominant voting share. Further details and visualizations can be found in Figures OA.1 to OA.4 in the online appendix 1. This implies that the Maker governance polls are characterized by different coalitions taking charge in a rotational manner. As a result, the majority of polls could be determined by a single voter coalition.

**[Figure 4 here]**

We then calculate the group cohesion of the identified voter coalitions within MakerDAO. The figure below illustrates that the group cohesion of all coalitions is generally high. However, the presence of a low minimum of AI indicates the existence of opinion differences within certain voting polls. Specifically, a low AI for a coalition in a particular poll suggests that the votes from that coalition are distributed among different options rather than concentrated on a single choice. This reflects a level of diversity or divergence of opinions within the coalition during that specific voting event.

**[Figure 5 here]**

Maker governance polls can be categorized based on their content and purpose. To enhance the understanding of the importance of these polls, the Maker Governance Portal labels each poll accordingly. The table below provides information on the number of different types of governance polls in which the three identified coalitions participated. Among these categories, the 'risk parameter' polls are particularly significant as they typically involve key parameters, such as the interest rates of

DAI loans. These polls directly impact the risk profile and overall stability of the Maker protocol. Consequently, the 'risk parameter' category holds considerable importance within the governance framework. Additionally, there are no significant differences in the participation of the various poll categories by the voter coalitions. Both voter coalition 1 and voter coalition 2 show relatively similar levels of participation in MIP and Greenlight polls. This implies that the coalitions are actively engaged in these types of polls and have a comparable level of influence and interest in shaping the decision-making process related to these polls.

[Table 3 here]

The last interesting point is to examine the internal structure of voter coalitions. Beside voting share and group cohesion, the identities of voters in different voter coalitions are of particular interest. Similarly with the principles of corporate finance around diverse boards (Bernile et al., 2018; Giannetti and Zhao, 2019), we attempt to reveal more information on the influences of diverse voters within the voter coalitions.

In order to provide a more comprehensive description of voters within MakerDAO, we have collected voters' Ethereum Name Service (ENS) names from the Maker Governance Portal. ENS serves as a unique identifier for blockchain addresses. Subsequently, we have searched for these ENS names on Twitter, as some blockchain users may use their ENS names as their Twitter handles. While most blockchain users tend to prefer anonymity, ENS owners and Twitter users may have a more public presence.

By analyzing voters' historical transactions, we can assign labels to describe their behavior, including activities such as decentralized exchange (DEX) trading, liquidity providing, and non-fungible token (NFT) trading. We also consider whether MakerDAO voters are considered "whales," which refers to entities holding a significant amount of tokens. Leveraging data from watchers.pro, we can examine voters' historical activities and assign the appropriate labels.

The table below provides an initial overview of the composition of the voter coalitions. Coalition 1 comprises the largest number of known users, indicating that the identities of these voters are more publicly transparent. Furthermore, there are 23 MakerDAO delegates within coalition 1, which enhances the influence of this coalition. Known users and delegates can utilize social media platforms like Twitter and the MakerDAO forum to influence the voting behavior of others. On the other hand, voters within coalitions 2 and 3 remain anonymous, as they do not have ENS names or known Twitter accounts. However, when it comes to the trading activities of voters, limited information about coalitions 2 and 3 is available. It's important to note that since an individual can have multiple addresses on the blockchain, some voters may have additional unknown addresses for their trading activities.

[Table 4 here]

The table below provides further details about the known voters and MakerDAO delegates within coalition 1. It is worth noting that several influential entities have been identified. For instance, Andreessen Horowitz (a16z), one of the prominent venture capital firms, has been detected and has participated in 6 polls. We also find some voters relevant to prestigious universities, such as 'Penn Blockchain' and 'Blockchain@Columbia'. Additionally, there are voters associated with prestigious universities, such as 'Penn Blockchain' and 'Blockchain@Columbia'. Upon reviewing their Twitter accounts, it becomes apparent that these two voters represent student organizations focused on blockchain and crypto-curious students. Although we may not have information on how they have acquired significant voting power (i.e., a large number of MKR tokens), the presence of voters affiliated with higher education institutions suggests that students from elite universities can exert influence on decentralized finance (DeFi) by actively participating in DAO governance. Furthermore, coalition 1 comprises crypto-native enterprises like Gauntlet[4] and Flipside[5], as well as influencers within the cryptocurrency industry such as Hasu[6] and Chris Blec[7]. The diversity in coalition 1 makes it more intriguing to examine how this coalition affects Maker protocol.

**[Table 5 here]**

## 4. The influence of voter coalitions

This section investigates how voter coalitions influence Maker protocol. In corporate finance, shareholders will join forces with other to reduce uncertainty (Hogg, 2000), so strong coalitions can emerge (Sauerwald and Peng, 2013). For example, large shareholders can form coalitions to extract private benefits (Bennedsen and Wolfenzon, 2000), while minority shareholders will collaborate to protect their own interests (Zwiebel, 1995; Bennedsen and Wolfenzon, 2000). Furthermore, if shareholder coalitions have sufficient voting rights to influence decision making, interest conflicts should be observed (Sauerwald and Peng, 2013; Marquardt et al., 2018), especially when it comes to controversial proposals (Ginzburg et al., 2022). However, the impact of conflicts between shareholder coalitions are not well studied (Dyck and Zingales, 2004).

Given that voting history of MakerDAO is transparent, we are able to investigate how voter coalitions affect Maker protocol. Here, we focus on three specific factors of Maker protocol, including DAI volatility, daily revenue of Maker protocol, and new users. To estimate regressions, two poll-level measurements, i.e., voting share and AI of voter coalitions, are transferred to daily measurements by taking weighted average, where weights are total votes of polls. The descriptive statistics of voting share and AI on daily basis are given in appendix 2. The following analysis is based on Poll 413 – Poll 838, where we do not find voting behavior of coalition 3[8].

---

[4] https://gauntlet.network/
[5] https://flipsidecrypto.xyz/
[6] https://twitter.com/hasufl?s=20
[7] https://twitter.com/ChrisBlec?s=20

**4.1 DAI Volatility**

We are interested in examining the impact of voter coalitions on DAI volatility. The primary objective of the DAI stablecoin is to maintain price stability, with 1 DAI intended to be pegged to 1 US dollar. High volatility is generally considered unfavorable for stablecoins (Gans, 2023; Liu, Makarov and Schoar, 2023). In the field of corporate finance, centralized governance structures have been linked to performance volatility, such as increased volatility in stock returns (Giannetti and Zhao, 2019; Tran and Turkiela, 2020). In the context of DAO governance, we anticipate that centralized voting power held by voter coalitions may have similar effects. In light of this, we estimate the following regression models:

$$\begin{aligned}
Dai\ volatility_t = & \beta_0 + \beta_1 voting\ share\ 1_t + \beta_2 voting\ share\ 2_t \\
& + \beta_3 AI\ 1_t + \beta_4 AI\ 2_t + \beta_5 voting\ share\ 1_t \times AI\ 1_t + \beta_6 voting\ share\ 2_t \times AI\ 2_t \\
& + \beta_7 \Delta ETH_t + \beta_8 \Delta RWA_t + \beta_9 Dai\ volume_t + \beta_{10} Mkr\ return_t + \beta_{11} ETH\ volatility_t \\
& + \varepsilon_t (5)
\end{aligned}$$

In addition to voting share and AI, we also incorporate two interaction terms, namely *voting share 1 × AI 1* and *voting share 2 × AI 2*, as the influence of a coalition depends on both voting power and group cohesion. Furthermore, we consider several other explanatory variables that capture three categories of influential factors for the Maker protocol. Detailed definitions of these variables can be found in Table A.3 in Appendix 3.

*ΔETH* and *ΔRWA* represent the changes in collateral assets locked in the Maker protocol. Typically, Ether (ETH) serves as the primary collateral asset since it is the native cryptocurrency on the Ethereum blockchain. However, Real World Assets (RWA) have gained importance as collateral in the Maker protocol recently, and certain governance polls revolve around the acceptance of specific RWAs as collateral. *Dai volume* and *Mkr return* are related to the performance of the Maker protocol. Volume is a key metric for financial products, and higher trading volume indicates a positive market performance for DAI. MKR, as the governance token in the Maker protocol, is akin to stocks in corporate finance. Therefore, a higher daily return of MKR reflects positive expectations for the Maker protocol. Lastly, *ETH volatility* is included in the regression models as it provides insights into the state of DeFi markets based on the Ethereum blockchain. We calculate the variance inflation factor

---
[8] Given a coalition, we let AI equal to one when none of the voters participated in a governance poll since all voters in the coalition in fact choose 'abstain'. After transferring the poll-level AI to the series on the daily basis, we estimate the regression models in the following subsections. In the following regression models, we do not include variables relevant to coalition 3, given that they did not participate in the polls.

(VIF) for these variables and find no evidence of multicollinearity.

The empirical results reveal several interesting findings. Firstly, a higher voting share 2 is associated with a decrease in DAI volatility. This implies that the presence of the smaller voter coalition contributes to the price stability of DAI. However, we do not find evidence of coalition 1's influence on DAI volatility, except in column (1). These findings expand upon the discussion in Bernile et al. (2018), which suggests that centralization can lead to higher volatility. In the case of cryptocurrencies, centralized governance power may not directly cause volatility, but minority voters, particularly when they form a coalition, can impact the price dynamics of stablecoins.

Additionally, in columns (5) to (8), DAI volume demonstrates positive effects on DAI volatility, suggesting a trade-off between volume and volatility for stablecoins. This finding aligns with research on return predictors in the cryptocurrency market (Liu, Tsyvinski, and Wu, 2022; Șoiman, Duma, and Jimenez-Garces, 2023). Interestingly, when ETH exhibits higher volatility, DAI tends to exhibit better stability. This further contributes to the ongoing discussion regarding the interconnections between cryptocurrency returns (Guo, Härdle and Tao, 2022; Șoiman, Duma and Jimenez-Garces, 2023).

[Table 6 here]

**4.2 Total revenue of Maker protocol**

Maker protocol plays a role similar to that of banks within the DeFi ecosystem, with a significant portion of its total revenue derived from DAI loans. Since various aspects of DAI loans, including interest rates and acceptable collateral assets, are determined through on-chain governance, the decisions made by MakerDAO directly impact loan volumes and, consequently, the revenue generated by the Maker protocol. To investigate the relationship between protocol revenue and the decision-making power of voter coalitions, we estimate the following regression model:

$$\begin{aligned}
Daily\ revenue_t = &\ \beta_0 + \beta_1 voting\ share\ 1_t + \beta_2 voting\ share\ 2_t \\
&+ \beta_3 AI\ 1_t + \beta_4 AI\ 2_t + \beta_5 voting\ share\ 1_t \times AI\ 1_t + \beta_6 voting\ share\ 2_t \times AI\ 2_t \\
&+ \beta_7 \Delta ETH_t + \beta_8 \Delta RWA_t + \beta_9 Dai\ volume_t + \beta_{10} Mkr\ return_t + \beta_{11} ETH\ volatility_t \\
&+ \varepsilon_t \quad (6)
\end{aligned}$$

The dependent variable in regression (6) is the daily revenue of the Maker protocol, measured in USD. The explanatory variables align with those used in regression (5), and the results are presented in the table below. The voting share of the smaller coalition, indicated by a higher *voting share 2*, exhibits a positive influence on the daily revenue of the Maker protocol. This positive effect remains significant even after controlling for AI (see column (7)). Conversely, *voting share 1* has a negative impact in columns (1) and (5). These findings suggest that the dominant voter coalition can potentially undermine the health of DeFi protocols, as daily revenue is vital for their long-term growth. However, if a small voter coalition exists as an opposing force, DeFi protocols may experience increased revenue when the smaller coalition gains more voting power. Additionally, we observe a positive relationship between DAI volume and the daily revenue of the Maker protocol, which aligns with common sense.

[Table 7 here]

The aforementioned findings contribute to the broader discussion on corporate governance. Previous research, such as that conducted by Shleifer and Vishny (1997) and Maury and Pajuste (2005), has shown that active large shareholders may or may not bring benefits, and their voting strategies are often tied to their own private interests. Furthermore, if a minority voting coalition emerges victorious, their decisions may deviate from previous ones (Garlappi et al., 2017). In the context of the Maker protocol, it is possible that some voters may attempt to manipulate governance outcomes for their personal gains. In such cases, it is likely that we would observe a decrease in daily revenue when selfish minority voters have a higher likelihood of winning a governance poll. Similar issues of self-interest within corporations have been discussed by Berglof and Burkart (2003). Shareholders have the ability to divert corporate resources for personal benefits, and this profit-seeking behavior can influence their decision-making in corporate governance (Hamdani and Yafeh, 2013; Cvijanović et al., 2016). Intuitively, DeFi governance is not exempt from conflicts of interest, and voter coalitions, along with their decisions, can have either positive or negative consequences.

**4.3 New users of Maker protocol**

Network adoption plays a pivotal role in the success of decentralized digital platforms. Cong, Li, and Wang (2020) and Xiong and Sockin (2023) have demonstrated how user adoption influences platform growth. In the case of DeFi, attracting new users is equally crucial and (2023) discuss network effect in DeFi market. However, it remains uncertain whether voter coalitions can impact the growth of Maker users. To investigate this, we utilize the datasets for users of the Maker protocol from Dune.xyz. By calculating the number of new users on a daily basis, we can estimate the following regression model:

$$\begin{aligned} New\ MakerDAO_t &= \beta_0 + \beta_1 voting\ share\ 1_t + \beta_2 voting\ share\ 2_t \\ &+ \beta_3 AI\ 1_t + \beta_4 AI\ 2_t + \beta_5 voting\ share\ 1_t \times AI\ 1_t + \beta_6 voting\ share\ 2_t \times AI\ 2_t \\ &+ \beta_7 \Delta ETH_t + \beta_8 \Delta RWA_t + \beta_9 Dai\ volume_t + \beta_{10} Mkr\ return_t + \beta_{11} ETH\ volatility_t \\ &+ \varepsilon_t \quad (7) \end{aligned}$$

The table below presents some noteworthy findings. We observe that *voting share 2* is positively correlated with the number of new daily users, while higher *voting share 1* has a negative impact on user growth in certain instances (e.g., columns (1) and (5)). This finding suggests that a large voter coalition may undermine the overall health of the Maker protocol, whereas smaller coalitions can be beneficial, which aligns with the results from the previous subsections. Additionally, we find that the group cohesion within voter coalitions plays a role. For instance, a higher *AI 2*, indicating a more cohesive coalition 2, can have a negative effect on user adoption (e.g., columns (6) and (8)), and the intersection term *voting share 2 × AI 2* exhibits a positive coefficient in column (8). These results suggest that a cohesive smaller voter coalition can also impede user adoption, but the negative influence diminishes when the smaller coalition gains more voting power.

[Table 8 here]

A potential explanation is the uncertainty surrounding group decisions. Uncertainty can arise from the diverse beliefs held by decision makers (Ellsberg, 1961), and theoretical research, such as that conducted by Hayashi and Lombardi (2018), further explores decision-making under conditions of uncertainty. In the context of corporate governance, minority shareholders can introduce an element of uncertainty. If minority decision-makers succeed in the governance process, group decisions may exhibit inconsistencies (Garlappi et al., 2017). In the case of MakerDAO, smaller coalitions play a similar role, and the decisions made by these minority coalitions may deviate from previous decisions. Therefore, the presence of small voter coalitions can be a concern for potential users of the Maker protocol. Although small voter coalitions can partially mitigate governance centralization, the adoption of DeFi in the network may be influenced by the "partisan fight" between these voter coalitions.

**4.4 Following the money: DAI**

As DAI is primary cryptocurrency issued by Maker protocol, the decision-making process of MakerDAO probably has influence on the flows of DAI. In this section, we consider five different destinations of DAI financial flows, including *Centralized Exchanges (CeFi)*, *Decentralized Exchanges (DEXes)*, *Lending Protocols (LPs)*, *External Owned Accounts (EOAs)*, and *Bridges*. Simply, EOAs are accounts controlled by people (instead of codes), and bridges enable cross-chain transactions. For more details about these on-chain applications, we refer readers to online appendix 1.

The decision-making process within MakerDAO, through governance, can potentially influence the distribution of DAI across these destinations. Key decisions related to DAI, such as interest rates and collateral assets, are determined through the governance process. Therefore, it is reasonable to hypothesize that voter coalitions within MakerDAO can impact the flows of DAI. To that end, we

estimate the regressions below:

$$\begin{aligned}Dai\ transferred_{i,t} &= \beta_0 + \beta_1 voting\ share\ 1_t + \beta_2 voting\ share\ 2_t \\&+ \beta_3 AI\ 1_t + \beta_4 AI\ 2_t + \beta_5 voting\ share\ 1_t \times AI\ 1_t + \beta_6 voting\ share\ 2_t \times AI\ 2_t \\&+ \beta_7 \Delta ETH_t + \beta_8 \Delta RWA_t + \beta_9 Dai\ volume_t + \beta_{10} Mkr\ return_t + \beta_{11} ETH\ volatility_t \\&+ \varepsilon_t \quad (8)\end{aligned}$$

Where:

- $Dai\ transferred_i = \{CeFi, DEX, LP, EOA, Bridge\}$

The table below presents the results. For the case of voting share, we only observe the positive effect of *voting share 2* on DAI flows to CeFi. Considering together with the positive coefficient of *AI 2* in column (1), voter coalition 2 can bring CeFi more DAI stablecoins. But voter coalition 1's cohesion will refrain DAI flows to CeFi, implying that the two coalitions have dissimilar interests. In column (4), we again observe the opposite effects of the two coalitions on DAI flows transferred to EOA. Intuitively, certain issues related to DAI (e.g., borrow rates) are decided by Maker governance, where voter coalitions are main participants. Our findings suggest that the decisions made by voter coalitions indirectly affect the financial flows of DAI to different on-chain applications, implying the dissimilar private interest of the coalitions.

**[Table 9 here]**

## 5. Conclusion

In this paper, we focus on the presence of voter coalitions in MakerDAO and highlight their impact on Maker protocol. In fact, it is found that three voter coalitions exist in MakerDAO after applying K-means to voting record of governance polls, and these voter coalitions have varying effects on Maker protocol, implying their dissimilar interests. The largest coalition, coalition 1, is found to have a positive relationship with DAI volatility, indicating that its influence may undermine the health of DAI as a stablecoin. On the other hand, smaller coalitions are associated with price stability of DAI, suggesting that they play a role in maintaining the peg to the US dollar. Moreover, coalition 1's voting share can decrease daily revenue of Maker protocol, while a small coalition with more voting power can bring in more protocol revenue. This suggests that voter coalitions and their voting behavior can have an impact on the financial performance of the underlying protocol.

Group cohesion, which reflects the level of agreement and unity within a coalition, is also important. When the large coalition is more cohesive, or the small coalition is less cohesive, DAI will be more stable. This can be attributed to the fact that cohesive coalitions are more likely to have predictable voting results, leading to increased certainty and stability in the protocol. Besides, we also observe relationship between financial flows of DAI and group cohesion of voter coalitions, suggesting that the coalitions' voting choices are relevant to their trading activities of DAI stablecoins.

This paper extends discussion in corporate governance to decentralized governance, which is a heat topic in the era of DeFi. For now, DAO is the most popular solution to decentralized governance, however, empirical studies (Sun, Stasinakis and Sermpinis, 2022; Fritsch, Müller and Wattenhofer, 2022) reveal centralized distribution of voting power in DAO. Compared with these studies, our paper further reveals collaboration of DAO voters and investigates how voter coalitions drives underlying DeFi. In corporate governance, shareholder coalitions are common, especially when some shareholders share similar goals (Black, 1990; Tajfel and Turner, 2004), and such coalitions can influence corporate decisions (Zwiebel, 1995). Though researchers realize the importance of voter coalitions, it is not very easy to present empirical research because of data limitations (Adam, Hermalin and Weisbach, 2010). Fortunately, benefiting from blockchain technology, decision-making process in DAO is transparent and precise, and anyone can easily retrieve the full voting history of any DAO. Therefore, DAO allows us to better explore the relationship between governance and performance of a financial system, and our paper can serve as a best example of research on on-chain governance.

Our findings can further contribute to arguments about shareholder behavior in corporate governance. Usually, shareholders make decisions based on their private benefits, and voting results depend on interactions among shareholders (Bennedsen and Wolfenzon, 2000). To illustrate dissimilar interests of voter coalitions, we choose cash flows of DAI as a proxy and track how coalitions drive DAI to different on-chain financial applications. Interestingly, coalitions will increase DAI flows to on-chain applications very differently. For example, coalition 0 and coalition 2 prefer DAI flows to LPs and EOAs, while voting coalition 3 is their common 'enemy'. Such observations are consistent with Meffert and Gschwend (2007) and May and Armstrong (2010), where several voter coalitions may work together for certain goals. Further study can attempt to investigate more interactions between voter coalitions in DAO and examine if such interactions can affect the underlying DeFi protocol. Moreover, if DAO voters' private benefits can be revealed, we will be able to know more about why coalitions are formed and show how private benefits drive DeFi protocols.

Although our findings appear conceptually and empirically robust, they should be interpreted with their limitations in mind. First, we cluster voters based on their voting records, because there is no clear coalition membership in DAO. But we should realize that DAO community will not reveal their vulnerability by committing the potential coalitions of voters. Second, it is hard to identify the opinion leaders in voter coalitions because of anonymity of blockchain. Though not all voters can be matched with an entity in real life, some voters with publicly known identities, e.g., ENS names and Twitter, may be opinion leaders in MakerDAO. Further studies may attempt to reveal more information of

DAO voters, and the leaders in DAO will be an intriguing research topic. Third, we do not investigate why voter coalitions' voting share and group cohesion change. Voters, beside voting, can have more complex activities using the underlying DeFi, and their voting behavior can be influenced by all their financial activities. Therefore, motivations for participating in DAO should be better studied. Finally, assuming that voter coalitions exist, voters can turn to different voter coalitions. Therefore, the influence of multicoalition democracy on DAO and its underlying protocol should be more complicated than the discussion in this paper.

**Tables**

**Table 1. Descriptive statistics of Maker governance polls**

|  | Total votes | Total voters | Breakdown votes | Breakdown ratio | Vote share of the largest voter coalition |
|---|---|---|---|---|---|
| **Mean** | 47934.36 | 25.97 | 40492.13 | 0.88 | 0.74 |
| **Median** | 37477.47 | 22 | 34011.03 | 0.98 | 0.72 |
| **Maximum** | 293911.44 | 158 | 176846.86 | 1 | 1.00 |
| **Minimum** | 259.74 | 5 | 232.80 | 0.35 | 0.37 |
| **Std** | 33940.98 | 15.92 | 26878.17 | 0.17 | 0.16 |

Note: This table shows the descriptive statistics of Maker governance polls. For each poll, we compute the total votes and the number of total voters. 'Breakdown votes' refers to the votes of the winning option, and 'breakdown ratio' is breakdown votes divided by total votes. Finally, we calculate the largest voter coalition's voting share, which equals to the votes casted by coalition 0 divided by the total votes.

**Table 2. Descriptive statistics of voter coalitions in MakerDAO**

|  | Number of voters | Involved polls | Total votes | Since |
|---|---|---|---|---|
| **Voter coalition 1** | 153 | 771 | 28563027.39 | 2017-12-18 |
| **Voter coalition 2** | 14 | 650 | 7257261.53 | 2019-04-11 |
| **Voter coalition 3** | 5 | 149 | 2364382.05 | 2020-04-29 |

Notes: This table presents information about three voter coalitions detected by K-means algorithm. Voter coalition 1 has the highest number of voters and participated in more governance polls compared to the two smaller coalitions. The column 'since' is the date when the voters in a coalition first initiated a transaction on Ethereum blockchain.

**Table 3. Voting participation in different categories of governance polls**

|  | Voter coalition 1 | Voter coalition 2 | Voter coalition 3 |
|---|---|---|---|
| **Risk parameter** | 297 | 262 | 52 |
| **MIP** | 181 | 112 | 29 |
| **Greenlight** | 173 | 152 | 42 |
| **Ratification poll** | 103 | 33 | 1 |
| **Inclusion poll** | 70 | 70 | 25 |
| **Collateral onboarding** | 63 | 55 | 13 |
| **Total participated polls** | 771 | 650 | 149 |

Note: This table presents the number of different categories of governance polls (Poll 16 – Poll 838) that the three voter coalitions participated in. More details about the categories of governance polls are given in Table A.1 in appendix 1.

**Table 4. Internal structure of voter coalitions**

|  | Voter coalition 1 | Voter coalition 2 | Voter coalition 3 |
|---|---|---|---|
| **ENS owner** | 30 | 0 | 0 |
| **Twitter user** | 16 | 0 | 0 |
| **DEX trader** | 67 | 7 | 0 |
| **Liquidity provider** | 30 | 2 | 0 |
| **NFT trader** | 27 | 5 | 0 |
| **Whale** | 20 | 1 | 0 |
| **Delegate** | 23 | 0 | 0 |
| **Total voters** | 153 | 14 | 5 |

Note: This table describes the internal structure of voter coalitions in MakerDAO. For each voter coalition, we list the number of voters in different categories.

**Table 5. Voters with known identities and MakerDAO delegates in coalition 1**

| Address | ENS name | Twitter | Delegate | Total votes | Involved polls |
|---|---|---|---|---|---|
| 0xaf8aa6846539033eaf0c3ca4c9c7373e370e039b | Flip Flop Flap Delegate LLC | CruzerDefi | 1 | 4555078.4 | 225 |
| 0xb21e535fb349e4ef0520318acfe589e174b0126b | schuppi | schuppi | 1 | 2735238.67 | 228 |
| 0x845b36e1e4f41a361dd711bda8ea239bf191fe95 | Feedblack Loops LLC | | 1 | 1928371.78 | 212 |
| 0xad2fda5f6ce305d2ced380fdfa791b6a26e7f281 | Field Technologies, Inc. | ImperiumPaper | 1 | 1271752.98 | 67 |
| 0x22d5294a23d49294bf11d9db8beda36e104ad9b3 | MakerMan | | | 1071135.69 | 217 |
| 0x45127ec92b58c3a89e89f63553073adcaf2f1f5f | monetsupply | monetsupply | 1 | 943768.66 | 199 |
| 0x00daec2c2a6a3fcc66b02e38b7e56dcdfa9347a1 | | | 1 | 936786.68 | 45 |
| 0x4d3ac33ab1dd7b0f352b8e590fe8b62c4c39ead5 | ACREinvest | ACREinvest | 1 | 548753.02 | 130 |
| 0xb0b829a6aae0f7e59b43391b2c8a1cfd0c801c8c | gauntlet | gauntletnetwork | 1 | 468000 | 102 |
| 0xcdb792c14391f7115ba77a7cd27f724fc9ea2091 | JustinCase | | 1 | 463620.37 | 161 |
| 0x74971f1be0afd1bb820668abfe411d164f17b53c | | | 1 | 403948 | 12 |
| 0xefcc3401739427eb0491cc27c7baa06817c7dfdb | | | 1 | 394429.2 | 69 |
| 0xafaff1a605c373b43727136c995d21a7fcd08989 | Hasu | hasufl | 1 | 390896.15 | 43 |
| 0xf60d7a62c98f65480725255e831de531efe3fe14 | GFX Labs | labsGFX | 1 | 274272.97 | 159 |
| 0x8804d391472126da56b9a560aef6c6d5aaa7607b | Doo | DooWanNam | 1 | 213828.87 | 98 |
| 0x05e793ce0c6027323ac150f6d45c2344d28b6019 | a16z | a16z | | 156960 | 6 |
| 0x68b216e9fc96a7b98b5c0028ff72e4c39c5c5a61 | | | 1 | 136909 | 29 |
| 0x84b05b0a30b6ae620f393d1037f217e607ad1b96 | Flipside Crypto | Flipsidecrypto | 1 | 116384.68 | 77 |
| 0x2c3b917cceaf41503145ceb4b37c8623d862c4cd | | | 1 | 102000 | 59 |
| 0x2c511d932c5a6fe4071262d49bfc018cfbaaa1f5 | Chris Blec | ChrisBlec | 1 | 91578.52 | 25 |
| 0x7ddb50a5b15aea7e7cf9ac8e55a7f9fd9d05ecc6 | Penn Blockchain | PennBlockchain | 1 | 73911.4 | 67 |
| 0xb8df77c3bd57761bd0c55d2f873d3aa89b3da8b7 | Blockchain@Columbia | BlockchainatCU | 1 | 22000 | 22 |
| 0x14a4ed2000ca405452c140e21c10b3536c1a98e4 | | | 1 | 15566.5 | 234 |
| 0xaa19f47e6acb02df88efa9f023f2a38412069902 | mhonkasalo & teemulau | mhonkasalo;teemulau | 1 | 8023.87 | 8023 |
| 0x4e314eba76c3062140ad196e4ffd34485e33c5f5 | Governance House | | 1 | 7007 | 7 |
| 0xe84adc0964ee34ce0319df3418636ed6a4117b97 | justneedtogetthroughthisweek.eth | | | 6569.84 | 143 |
| 0x14341f81df14ca86e1420ec9e6abd343fb1c5bfc | tylersorensen.eth | | | 6106.56 | 30 |
| 0x4f2fc90212e949ff4aa32def570744163671f22b | 00x.eth | 00x_eth | | 818.44 | 79 |
| 0x57db5d6aa783cf29af41330569d24957140fd3eb | dix-sept.eth | | | 736.85 | 122 |
| 0xa7bc2dc8d3ea8ef85faf48d560fa56835abcea88 | blockworm.eth | | | 620 | 10 |

Note: This table provides detailed information about voters with known identities and MakerDAO delegates in coalition 1.

ENS names are publicly available, and the column 'twitter' listed the twitter usernames if a voter is detected as a Twitter user. If a voter is a MakerDAO delegate, the value in column 'delegate' is 1. We find that some delegates do not disclose any public identities (i.e., ENS names and Twitter accounts).

**Table 6. DAI volatility and voter coalitions**

|  | (1) | (2) | (3) | (4) | (5) | (6) | (7) | (8) |
|---|---|---|---|---|---|---|---|---|
| **Voting share1** | **0.05*** | - | 0.02 | 0.33 | 0.03 | - | 0.00 | 0.18 |
|  | **(1.83)** |  | (0.49) | (0.78) | (1.41) |  | (0.13) | (0.44) |
| **Voting share2** | **-0.20*** | - | **-0.20*** | 0.07 | **-0.16*** | - | **-0.16*** | 0.06 |
|  | **(-3.28)** |  | **(-2.97)** | (0.11) | **(-2.66)** |  | **(-2.48)** | (0.09) |
| **AI1** | - | **-0.10*** | -0.09 | 0.09 | - | **-0.09*** | -0.09 | 0.01 |
|  |  | **(-1.98)** | (-1.37) | (0.35) |  | **(-1.76)** | (-1.41) | (0.04) |
| **AI2** | - | **0.16*** | 0.06 | 0.07 | - | 0.13 | 0.04 | 0.06 |
|  |  | **(1.76)** | (0.59) | (0.61) |  | (1.41) | (0.41) | (0.48) |
| Voting share1*AI1 | - | - | - | -0.31 | - | - | - | -0.18 |
|  |  |  |  | (-0.74) |  |  |  | (-0.42) |
| Voting share2*AI2 | - | - | - | -0.24 | - | - | - | -0.21 |
|  |  |  |  | (-0.35) |  |  |  | (0.32) |
| ΔETH | - | - | - | - | 0.03 | 0.01 | 0.03 | 0.03 |
|  |  |  |  |  | (0.25) | (0.09) | (0.23) | (0.22) |
| ΔRWA | - | - | - | - | 0.01 | 0.03 | 0.02 | 0.02 |
|  |  |  |  |  | (0.06) | (0.21) | (0.12) | (0.13) |
| **Dai volume** | - | - | - | - | **0.73*** | **0.73*** | **0.72*** | **0.72*** |
|  |  |  |  |  | **(6.06)** | **(6.06)** | **(6.03)** | **(6.01)** |
| Mkr return | - | - | - | - | -0.03 | -0.06 | -0.04 | -0.03 |
|  |  |  |  |  | (-0.35) | (-0.62) | (-0.37) | (-0.28) |
| **ETH volatility** | - | - | - | - | **-0.37*** | **-0.41*** | **-0.37*** | **-0.37*** |
|  |  |  |  |  | **(-4.51)** | **(-4.93)** | **(-4.52)** | **(-4.47)** |
| **Constant** | 0.07 | 0.10 | **0.20*** | 0.33 | **0.18**** | 0.17 | 0.24 | 0.12 |
|  | (0.71) | (1.15) | **(1.74)** | (0.78) | **(1.99)** | (1.24) | (1.59) | (0.39) |
| **N** | 533 | 533 | 533 | 533 | 533 | 533 | 533 | 533 |
| **Adj. R-sq** | 0.02 | 0.01 | 0.02 | 0.02 | 0.09 | 0.09 | 0.09 | 0.09 |

Note: This table reports the regression coefficients and standard t-statistics in the parentheses for the case of DAI volatility. Columns (1) – (4) present results for baseline regression models, where variables related to Maker protocol and Ethereum markets are not included. Columns (5) – (8) present results for regression models with variables related to Maker protocol and Ethereum markets. *, **, and *** denote significance levels at the 10%, 5%, and 1% levels based on the standard t-statistics. The definitions of the variables are given in Table A.3.

**Table 7. Daily revenue of Maker protocol and voter coalitions**

|  | (1) | (2) | (3) | (4) | (5) | (6) | (7) | (8) |
|---|---|---|---|---|---|---|---|---|
| **Voting share1** | **-0.07*** | - | -0.06 | -0.37 | **-0.07*** | - | -0.06 | -0.41 |
|  | **(-2.27)** |  | (-1.36) | (-0.76) | **(-2.31)** |  | (-1.40) | (-0.82) |
| **Voting share2** | **0.20*** | - | **0.22*** | 0.25 | **0.19*** | - | **0.21*** | 0.17 |
|  | **(2.82)** |  | **(2.83)** | (0.33) | **(2.64)** |  | **(2.65)** | (0.22) |
| AI1 | - | 0.08 | 0.02 | -0.17 | - | 0.08 | 0.02 | -0.19 |
|  |  | (1.27) | (0.24) | (-0.56) |  | (1.33) | (0.26) | (-0.62) |
| AI2 | - | -0.04 | 0.07 | 0.08 | - | -0.04 | 0.06 | 0.07 |
|  |  | (-0.36) | (0.56) | (0.59) |  | (-0.34) | (0.50) | (0.50) |
| Voting share1*AI1 | - | - | - | 0.33 | - | - | - | 0.36 |
|  |  |  |  | (0.65) |  |  |  | (0.71) |
| Voting share2*AI2 | - | - | - | -0.08 | - | - | - | -0.01 |
|  |  |  |  | (-0.10) |  |  |  | (-0.01) |
| ΔETH | - | - | - | - | 0.10 | 0.14 | 0.11 | 0.11 |
|  |  |  |  |  | (0.68) | (0.91) | (0.72) | (0.73) |
| ΔRWA | - | - | - | - | -0.01 | -0.03 | 0.00 | 0.00 |
|  |  |  |  |  | (-0.04) | (-0.21) | (-0.01) | (-0.02) |
| **Dai volume** | - | - | - | - | **0.26*** | 0.24 | **0.25*** | **0.25*** |
|  |  |  |  |  | **(1.74)** | (1.59) | **(1.71)** | **(1.73)** |
| Mkr return | - | - | - | - | -0.11 | -0.07 | -0.11 | -0.12 |
|  |  |  |  |  | (-0.93) | (-0.64) | (-0.95) | (-1.03) |
| **ETH volatility** | - | - | - | - | 0.14 | **0.19*** | 0.14 | 0.13 |

|  | | | | | (1.35) | **(1.83)** | (1.36) | (1.30) |
|---|---|---|---|---|---|---|---|---|
| **Constant** | **0.39\*\*\*** | **0.35\*\*\*** | **0.31\*\*** | 0.48 | **0.34\*\*\*** | 0.26 | 0.25 | 0.45 |
|  | **(35.35)** | **(3.33)** | **(2.22)** | (1.41) | **(2.97)** | (1.57) | (1.38) | (1.24) |
| **N** | 540 | 540 | 540 | 540 | 540 | 540 | 540 | 540 |
| **Adj. R-sq** | 0.02 | 0.02 | 0.02 | 0.01 | 0.02 | 0.01 | 0.02 | 0.02 |

Note: This table reports the regression coefficients and standard t-statistics in the parentheses for the case of daily revenue of Maker protocol. Columns (1) – (4) present results for baseline regression models, where variables related to Maker protocol and Ethereum markets are not included. Columns (5) – (8) present results for regression models with variables related to Maker protocol and Ethereum markets. \*, \*\*, and \*\*\* denote significance levels at the 10%, 5%, and 1% levels based on the standard t-statistics. The definitions of the variables are given in Table A.3.

**Table 8. New users of Maker protocol and voter coalitions**

|  | (1) | (2) | (3) | (4) | (5) | (6) | (7) | (8) |
|---|---|---|---|---|---|---|---|---|
| **Voting share1** | **-0.06\*\*\*** | - | -0.03 | 0.07 | **-0.05\*\*** | - | -0.02 | 0.17 |
|  | **(-2.87)** |  | (-1.07) | (0.43) | **(-2.36)** |  | (-0.66) | (1.10) |
| **Voting share2** | **0.35\*\*\*** | - | **0.34\*\*\*** | -0.76 | **0.27\*\*\*** | - | **0.27\*\*\*** | -0.80 |
|  | **(6.60)** |  | **(5.88)** | (-1.38) | **(5.39)** |  | **(4.80)** | (-1.55) |
| **AI1** | - | **0.11\*\*** | 0.09 | 0.15 | - | **0.10\*\*\*** | **0.10\*** | **0.21\*\*** |
|  |  | **(2.44)** | (1.56) | (1.38) |  | **(2.48)** | **(1.76)** | **(1.99)** |
| **AI2** | - | **-0.27\*\*\*** | -0.09 | **-0.21\*\*** | - | **-0.23\*\*\*** | -0.09 | **-0.21\*\*** |
|  |  | **(-3.19)** | (-0.98) | **(-2.01)** |  | **(-2.90)** | (-1.05) | **(-2.14)** |
| **Voting share1\*AI1** | - | - | - | -0.12 | - | - | - | -0.20 |
|  |  |  |  | (-0.74) |  |  |  | (-1.34) |
| **Voting share2\*AI2** | - | - | - | **1.20\*\*** | - | - | - | **1.16\*\*** |
|  |  |  |  | **(2.05)** |  |  |  | **(2.13)** |
| **ΔETH** | - | - | - | - | **0.20\*\*** | **0.23\*\*** | **0.20\*** | **0.21\*\*** |
|  |  |  |  |  | **(1.93)** | **(2.17)** | **(1.92)** | **(2.02)** |
| **ΔRWA** | - | - | - | - | 0.03 | -0.01 | 0.01 | 0.01 |
|  |  |  |  |  | (0.23) | (-0.10) | (0.11) | (0.05) |
| **Dai volume** | - | - | - | - | **-0.29\*\*\*** | **-0.29\*\*\*** | **-0.28\*\*\*** | **-0.28\*\*\*** |
|  |  |  |  |  | **(-2.78)** | **(-2.79)** | **(-2.73)** | **(-2.75)** |
| **Mkr return** | - | - | - | - | **0.13\*** | **0.18\*\*** | **0.14\*** | 0.13 |
|  |  |  |  |  | **(1.64)** | **(2.13)** | **(1.68)** | (1.56) |
| **ETH volatility** | - | - | - | - | **0.66\*\*\*** | **0.71\*\*\*** | **0.66\*\*\*** | **0.66\*\*\*** |
|  |  |  |  |  | **(9.21)** | **(9.91)** | **(9.24)** | **(9.36)** |
| **Constant** | **0.22\*\*\*** | **0.3\8\*\*\*** | **0.22\*\*** | **0.28\*** | -0.06 | 0.03 | -0.07 | -0.05 |
|  | **(27.05)** | **(4.69)** | **(2.09)** | **(1.84)** | (-0.71) | (0.27) | (-0.53) | (-0.31) |
| **N** | 539 | 539 | 539 | 539 | 539 | 539 | 539 | 539 |
| **Adj. R-sq** | 0.08 | 0.02 | 0.08 | 0.08 | 0.20 | 0.17 | 0.20 | 0.21 |

Note: This table reports the regression coefficients and standard t-statistics in the parentheses for the case of new users of Maker protocol. Columns (1) – (4) present results for baseline regression models, where variables related to Maker protocol and Ethereum markets are not included. Columns (5) – (8) present results for regression models with variables related to Maker protocol and Ethereum markets. \*, \*\*, and \*\*\* denote significance levels at the 10%, 5%, and 1% levels based on the standard t-statistics. The definitions of the variables are given in Table A.3.

**Table 9. DAI flows and voter coalitions**

|  | (1)<br>CeFi | (2)<br>DEX | (3)<br>LP | (4)<br>EOA | (5)<br>Bridge |
|---|---|---|---|---|---|
| Voting share1 | -0.29 | -0.15 | 0.23 | -0.21 | 0.20 |
|  | (-1.53) | (-0.58) | (0.91) | (-0.84) | (0.81) |
| **Voting share2** | **1.06\*** | 0.48 | 0.12 | 0.76 | 0.95 |
|  | **(1.65)** | (0.53) | (0.14) | (0.92) | (1.15) |
| **AI1** | **-0.30\*\*** | -0.24 | 0.25 | **-0.38\*\*** | 0.02 |
|  | **(-2.33)** | (-1.32) | (1.46) | **(-2.24)** | (0.10) |
| **AI2** | **0.23\*** | 0.18 | -0.17 | **0.31\*\*** | **0.28\*** |
|  | **(1.83)** | (1.06) | (-1.05) | **(1.95)** | **(1.74)** |
| Voting share1\*AI1 | 0.23 | 0.20 | -0.15 | 0.26 | -0.12 |
|  | (1.21) | (0.76) | (-0.61) | (1.07) | (-0.50) |
| **Voting share2\*AI2** | **-1.22\*** | -0.93 | -0.30 | -1.34 | -1.33 |
|  | **(-1.81)** | (-0.96) | (-0.34) | (-1.53) | (-1.53) |
| **ΔETH** | **-0.28\*\*** | **-0.45\*\*** | -0.06 | **-0.76\*\*\*** | **-0.91\*\*\*** |
|  | **(-2.13)** | **(-2.39)** | (-0.36) | **(-4.44)** | **(5.37)** |
| **ΔRWA** | 0.14 | 0.27 | 0.03 | 0.27 | **0.33\*** |

|  | (0.98) | (1.30) | (0.16) | (1.45) | **(1.78)** |
|---|---|---|---|---|---|
| **Dai volume** | **0.53\*\*\*** | **0.73\*\*\*** | -0.07 | **0.60\*\*\*** | 0.20 |
|  | **(4.14)** | **(4.05)** | (-0.41) | **(3.61)** | (1.23) |
| Mkr return | 0.02 | -0.17 | -0.20 | -0.15 | -0.09 |
|  | (0.21) | (-1.19) | (-1.52) | (-1.18) | (-0.70) |
| **ETH volatility** | **-0.38\*\*\*** | **-1.15\*\*\*** | **-0.41\*\*\*** | **-0.69\*\*\*** | **-0.90\*\*\*** |
|  | **(-4.32)** | **(-9.13)** | **(-3.47)** | **(-6.04)** | **(-7.92)** |
| **Constant** | **0.59\*\*\*** | **0.91\*\*\*** | **0.69\*\*\*** | **1.27\*\*\*** | **0.76\*\*\*** |
|  | **(2.85)** | **(3.08)** | **(2.48)** | **(4.70)** | **(2.85)** |
| N | 540 | 540 | 540 | 540 | 540 |
| Adj. R-sq | 0.06 | 0.18 | 0.03 | 0.18 | 0.17 |

Note: This table reports the regression coefficients and standard t-statistics in the parentheses for the case of new users of Maker protocol. Columns (1) – (5) present results for DAI flows transferred to different on-chain applications, where variables related to Maker protocol and Ethereum markets are also included. *, **, and *** denote significance levels at the 10%, 5%, and 1% levels based on the standard t-statistics. The definitions of the variables are given in Table A.3.

**Figures**
**Figure 1. Elbow method and silhouette score**

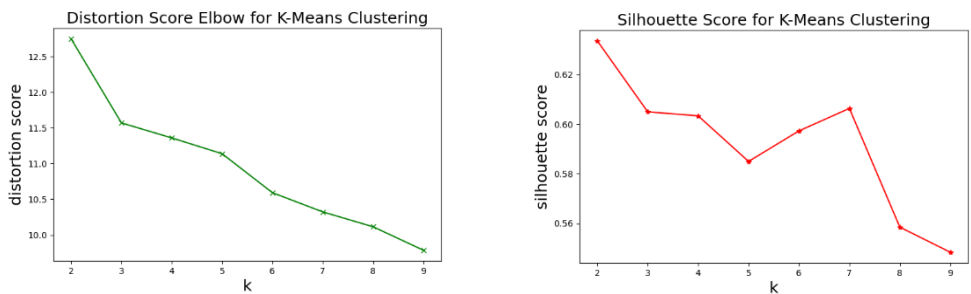

Note: This figure shows how we choose the optimal number of clusters when applying K-means algorithm. On the left, we use elbow method and compute distortion score, and this score measures the sum of squared distances from each point to its assigned center. Usually, distortion score could decrease rapidly at first then slowly flatten forming an "elbow" in a line graph, and we will choose the point where the score starts decreasing slowly as the optimal number of clusters. On the right, we calculate silhouette score, which measures how similar a data point is within-cluster compared to other clusters. Usually, we prefer choosing the number of clusters with the highest silhouette score. Combining with the line graph using elbow method, finally we choose 3 as the optimal number of clusters.

**Figure 2. Voters of Maker governance polls (Poll #16 – Poll #838)**

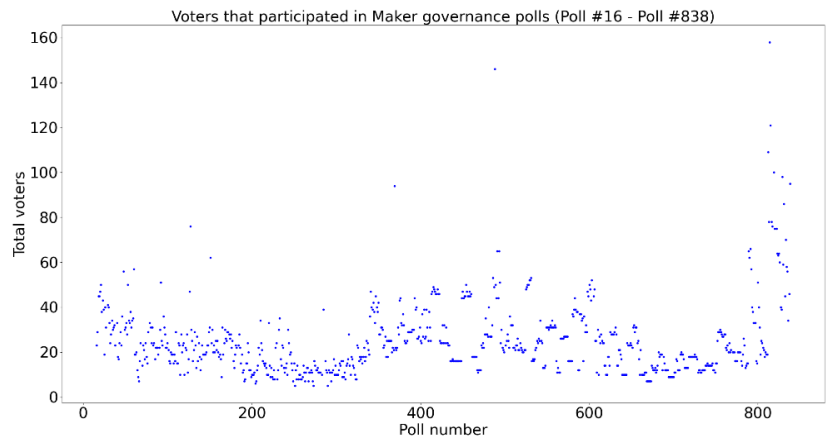

Note: This figure shows the number of voters in Maker governance polls (Poll #16 – Poll #838). In most polls, the number of voters will not be more than 60.

**Figure 3. Total votes of Maker governance polls (Poll #16 – Poll #838)**

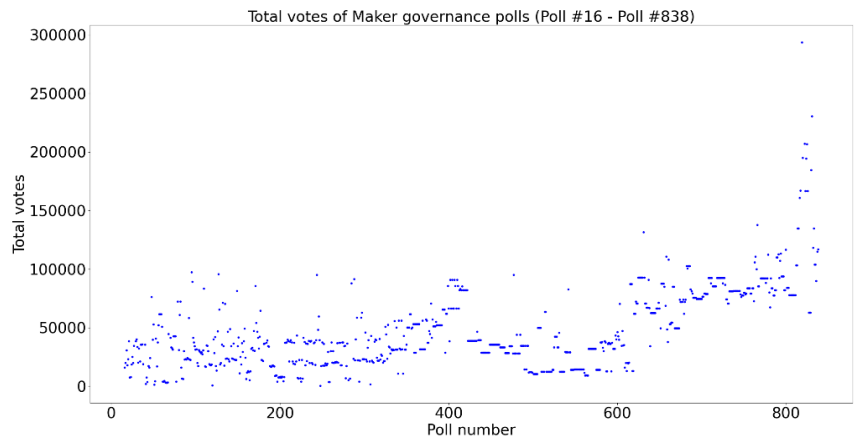

Note: This figure illustrates the total votes in Maker governance polls (Poll #16 – Poll #838). Overall, we observe that the total votes are on the increase, while most polls attract less than 100,000 votes.

**Figure 4. Voting share of three voter coalitions in MakerDAO**

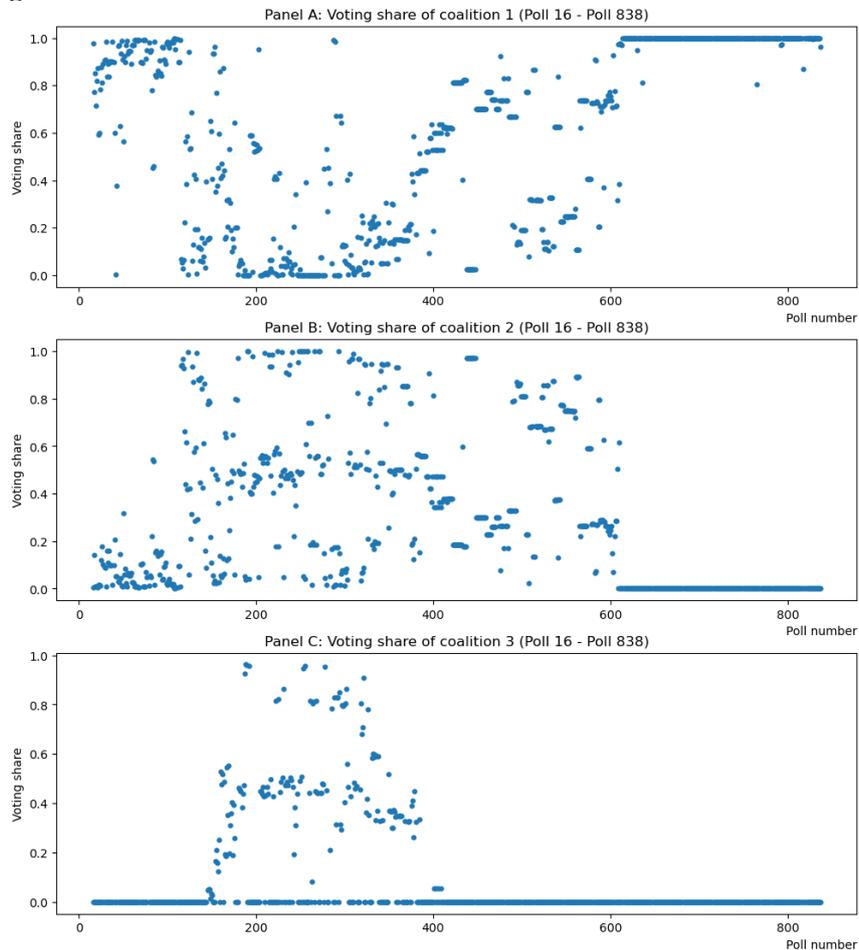

Note: This figure illustrates the voting share of three voter coalitions in Maker governance polls (Poll 16 – Poll 838). In some polls, the voting share of coalition 0 is close to 1, meaning that they have dominant decision-making power. In some other polls, the voting share of coalition 0 is low, while coalitions 2 and 3 contribute more votes, meaning that two smaller coalitions can win.

**Figure 5. Agreement Index (AI) of three voter coalitions in MakerDAO**

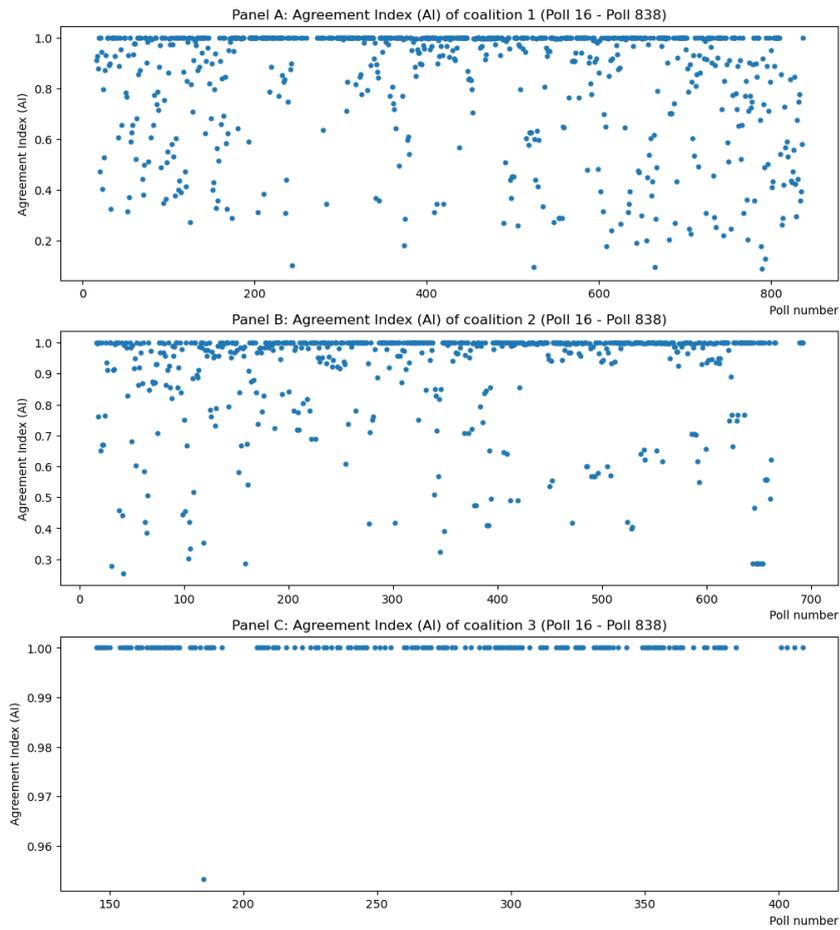

Note: This figure illustrates the Agreement Index (AI) of three voter coalitions in Maker governance polls (Poll 16 – Poll 838). In some polls, AI of coalition 0 is close to 1, meaning that they concentrate on the same option. In some other polls, AI is low, meaning that their voting power is dispersed.

# Appendices
## Appendix 1. Labels of Maker governance polls

**Table A.1. Labels of Maker governance polls**

|  | Number of polls | Total voters | Total votes |
|---|---|---|---|
| **Risk Parameter** | 297 | 6125 | 11620405.21 |
| **Ratification Poll** | 103 | 3180 | 9191601.36 |
| **Inclusion Poll** | 71 | 1514 | 2419753.25 |
| **Collateral Onboarding** | 63 | 1370 | 2864764.94 |
| **Collateral Offboarding** | 19 | 298 | 1381079.48 |
| **Greenlight** | 173 | 5549 | 7894197.78 |
| **Real World Asset** | 37 | 1051 | 2271616.02 |
| **Misc Governance** | 29 | 1108 | 1844854.05 |
| **Misc Funding** | 14 | 569 | 1480519.35 |
| **MakerDAO Open Market Committee** | 22 | 476 | 1294594.31 |
| **MIP** | 182 | 4800 | 11667401.86 |
| **Budget** | 61 | 1636 | 4511723.76 |
| **Oracle** | 42 | 761 | 1383196.90 |
| **System Surplus** | 10 | 263 | 721589.11 |
| **DAI Direct Deposit Module** | 10 | 215 | 820019.35 |
| **Multi-chain Bridge** | 5 | 126 | 385882.47 |
| **Technical** | 20 | 429 | 914487.67 |
| **Auction** | 23 | 421 | 715394.51 |
| **Delegates** | 5 | 53 | 338567.60 |
| **Peg Stability Module** | 14 | 252 | 643171.78 |
| **Core Unit Onboarding** | 29 | 899 | 1888838.96 |
| **Dai Savings Rate** | 28 | 662 | 959804.04 |
| **Black Thursday** | 4 | 172 | 265698.72 |
| **Multi-Collateral DAI Launch** | 5 | 165 | 192941.15 |
| **Prioritization Sentiment** | 2 | 55 | 54826.02 |

Note: There are several labels related to oracle. For convenience, we merge these labels into one category, namely 'oracle'.

## Appendix 2. Descriptive statistics of daily measurements

**Table A.2. Descriptive statistics**

| | Panel A: Voting share | | |
|---|---|---|---|
| | Voter coalition 1 | Voter coalition 2 | Voter coalition 3 |
| **Mean** | 0.15 | 0.03 | 0 |
| **Median** | 0 | 0 | 0 |
| **Maximum** | 1.00 | 0.97 | 0 |
| **Minimum** | 0 | 0 | 0 |
| **Std** | 0.14 | 0.14 | 0 |
| | Panel B: Agreement Index (AI) | | |
| | Voter coalition 1 | Voter coalition 2 | Voter coalition 3 |
| **Mean** | 0.80 | 0.91 | 0 |
| **Median** | 0.84 | 0.95 | 0 |
| **Maximum** | 1 | 1 | 0 |
| **Minimum** | 0.36 | 0.54 | 0 |
| **Std** | 0.19 | 0.12 | 0 |

Note: This table summarizes the descriptive statistics of voting shares and Agreement Index (AI) for three voter coalitions in MakerDAO, using the dataset for governance polls from Poll 413 to Poll 838. The statistics related to voter coalition 3 are zero, because we do not find voters in coalition 3 participated in Poll 413 – Poll 838.

## Appendix 3. Definitions of variables related to Maker protocol

**Table A.3. Definitions of variables**

| | Definitions |
|---|---|
| **ΔETH** | The changes of value of Ether (ETH) locked in Maker protocol as collateral |
| **ΔRWA** | The changes of value of Real World Asset (RWA) locked in Maker protocol as collateral |
| **Dai_volume** | Transaction volume (in USD) of DAI daily |
| **Mkr_return** | Daily return of Maker (MKR), which is the governance token in Maker protocol |
| **ETH volatility** | Daily return of Ether (ETH), which is the underlying cryptocurrency of Ethereum blockchain |

Note: This table presents definitions of explanatory variables in the regression models.

## Table A.4. Calculation of variables

| | Definitions |
|---|---|
| **Dai_volatility** | Assuming that the closing price of DAI on day $t$ is $P_t$, the daily volatility can be defined by $$V_t = \ln\left(\frac{P_t}{P_{t-1}}\right)$$ |
| **Daily revenue** | Assuming that $token_i, i = \{1, \ldots, n\}$, are locked in Maker protocol for lending, the variable 'daily revenue' can be calculated as $$\sum_{i=1}^{n} revenue_i$$ Where $revenue_i$ is the value in USD of revenue earned by the locked $token_i$. |
| **ΔETH** | Assuming that $ETH_t$ is the USD value of Ether (ETH) locked in Maker protocol for lending, the variable 'ΔETH' can be calculated as $$\Delta ETH_t = ETH_t - ETH_{t-1}$$ |
| **ΔRWA** | Assuming that $RWA_t$ is the USD value of Real World Assets (RWAs) locked in Maker protocol for lending, the variable 'ΔRWA' can be calculated as $$\Delta RWA_t = RWA_t - RWA_{t-1}$$ |
| **Mkr_return** | Assuming that the closing price of MKR on day $t$ is $P_t$, the daily volatility can be defined by $$V_t = \ln\left(\frac{P_t}{P_{t-1}}\right)$$ |
| **ETH_volatility** | Assuming that the closing price of ETH on day $t$ is $P_t$, the daily volatility can be defined by $$V_t = \ln\left(\frac{P_t}{P_{t-1}}\right)$$ |

Note: This table describes how to calculate variables included in the regression models.